\documentclass[conference,a4paper]{APSIPA2019}
\usepackage{multirow}
\usepackage[dvips]{graphicx}
\usepackage{epstopdf}
\usepackage{amsmath}
\usepackage[psamsfonts]{amssymb}
\usepackage{amsxtra}
\usepackage{threeparttable}
\usepackage{mathrsfs}

\renewcommand\arraystretch{1.2}

\begin{document}

\title{Triplet Based Embedding Distance and Similarity Learning for Text-independent Speaker Verification}

\author{%
\authorblockN{%
Zongze Ren, Zhiyong Chen and Shugong Xu
}
\authorblockA{%
Shanghai Institute for Advanced Communication and Data Science, Shanghai University, Shanghai, China \\
E-mail: \{zongzeren, bicbrv, shugong\}@shu.edu.cn}
}

\maketitle
\thispagestyle{empty}

\begin{abstract}
  Speaker embeddings become growing popular in the text-independent speaker verification task. 
In this paper, we propose two improvements during the training stage. The improvements are both based on triplet cause the training stage and the evaluation stage of the baseline x-vector system focus on different aims.
  Firstly, we introduce triplet loss for optimizing the Euclidean distances between embeddings while minimizing the multi-class cross entropy loss. Secondly, we design an embedding similarity measurement network for controlling the similarity between the two selected embeddings. We further jointly train the two new methods with the original network and achieve state-of-the-art. 
  The multi-task training synergies are shown with a 9\% reduction equal error rate (EER) and detected cost function (DCF) on the 2016 NIST Speaker Recognition Evaluation (SRE) Test Set.
  
\end{abstract}

\noindent\textbf{Index Terms}: speaker verification, deep neural network, similarity learning

\section{Introduction}
Speaker verification (SV) is a problem to give a decision that whether two utterances said by one person and can be defined as variable-length sequence classification task at utterance-level. 
The text-independent speaker verification (TI-SV) task is more challenging cause it does not have any lexicon or pronunciation constraints, the corpora also do not have transcript labels for training but only speaker information.

Previously, statistical models are extensively used for solving TI-SV problem, i-vector \cite{dehak2010front} based system is a representative of this type of approach and has achieved significant success in modelling speaker identity and channel variability in its space. Deep neural network (DNN) attracts more attention for TI-SV task recently. Frame acoustic sequences are extracted from raw audio signals and then stacked as the input of DNN. \cite{variani2014deep} uses several fully connected layers and the output number of the last layer corresponds to the number of speakers in the training processing. 
Cause predictions of frame-level \cite{variani2014deep} are not accurate enough, utterance-level representatives are considered by introducing several pooling operations. \cite{snyder2018x} uses statistical pooling of concatenating the average and standard deviation, called x-vector, which is recognized as a state-of-the-art solution. Other pooling methods such as self-attentive pooling \cite{zhu2018self}, attentive statistical pooling \cite{Okabe2018} and high-order statistical pooling \cite{you2019ustcspeech} also show their advantages. 

On the other hand, part of the knowledge involved in the training data is used to learn a classifier that will be ultimately thrown away in the verification tasks.
Triplet is a common way, FaceNet \cite{schroff2015facenet} firstly uses triplet loss in face-recognition task, the loss function minimizes the distance between an anchor and a positive while maximizes the distance between the anchor and a negative. \cite{zhang2017end, bredin2017tristounet, zhang2018text} use triplet loss in TI-SV task directly. However, systems only using single triplet loss are hard to converge in engineering. \cite{li2017deep} uses triplet loss to finetune the softmax pre-trained network, making it easy to train in reality.
After selecting a triplet, cosine similarity metric learning (CSML) \cite{novoselov2018triplet} is proposed to train a metric for back-end scoring to inplace tradtional LDA-PLDA. 
Furthermore, \cite{wan2018generalized} proposes a generalized end-to-end (GE2E) loss, which constructs a special batch to train speaker verification models more efficiently. And \cite{Yadav2018} trains the network architecture under the joint supervision of softmax loss and center loss.
Similarity task is a typical problem in natural language processing (NLP), there is no inherent ordering of the two sentences being compared. \cite{radford2018improving, devlin2018bert} concatenate the two utterance embeddings to contain both possible sentence orderings as the input of DNN.


Inspired by all these works, we put forward two methods based on triplet training while optimizing the multi-class classification target and test these two methods on Speaker Recognition Evaluation 2016 (SRE16) test set. Overall, proposed systems outperform the baseline system and here are our main contributions: (1) We choose a popular speaker verification network, x-vector as the baseline and implement Euclidean distances of a triplet, meaning that we minimize the distances between the anchor embedding and positive embedding while maximizes the distance between the anchor embedding and negative embedding, which reduces EER a lot. (2) We concatenate two embeddings of a triplet and feed them to a second DNN, named embedding similarity measurement network to measure the similarity between them, which makes a significant reduction of DCF by traditional LDA-PLDA backend scoring. (3) We jointly train the two methods with the original softmax-loss and achieve better performance. 
This idea of adding more constraints can extend to other fields.

The remainder of this paper is organized as follows. Section \uppercase\expandafter{\romannumeral2} reviews the x-vector structure of TI-SV tasks. In section \uppercase\expandafter{\romannumeral3}, we describe some details of the proposed improvements. The results of experiments and analysis are shown in Section \uppercase\expandafter{\romannumeral4}. Lastly, we give the conclusion in Section \uppercase\expandafter{\romannumeral5}.
\begin{figure*}[!h]
    \centering
    \includegraphics[scale=0.6]{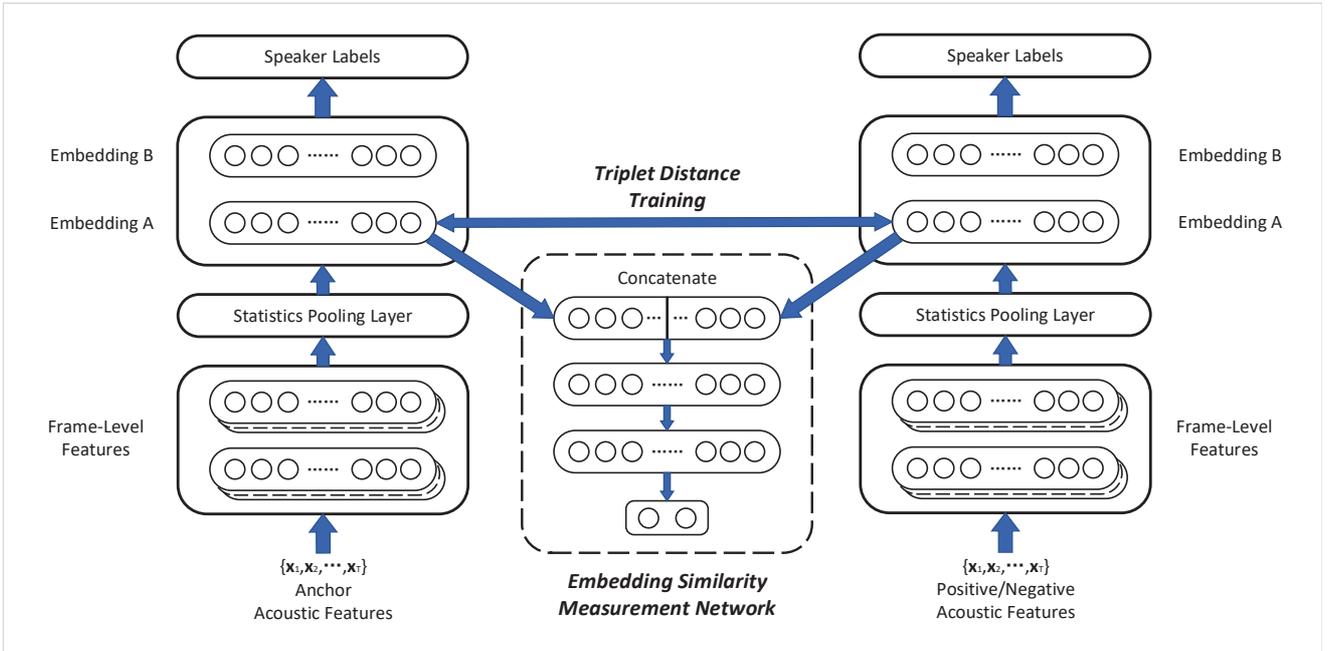}
    \caption{Proposed triplet distance training and embedding similarity measurement network with x-vector architecture}
    \label{fig:my_label}
\end{figure*}
\section{Baseline x-vector system}
\begin{table}
\caption{Baseline x-vector system architecture}\label{tab-xvector}\centering
\begin{tabular}{|c|c|c|c|}
\hline
Layer & Input-node & Input-dim & Output-dim\\
\hline
TDNN1 &	t-2,t-1,t,t+1,t+2 & 24*5 & 512\\
TDNN2 &	t-2,t,t+2 & 512*3 &	512\\
TDNN3 &	t-3,t,t+3 & 512*3 &	512\\
DNN4 & t & 512*3 &	512\\
DNN5 & t& 512 &	1500\\
\hline
stats pooling &T & 1500&3000\\
\hline
embedding a&T&3000&512 \\
embedding b&T&512&512 \\
softmax&T&512&num of speakers \\
\hline
\end{tabular}
\end{table}

In our work, speaker embedding is extracted by the x-vector architecture \cite{snyder2018x}, Table \ref{tab-xvector} shows the detail parameters. The whole system consists of three parts and the training process is end-to-end. Firstly, 23-dimensional MFCC features are feed to the network as frame-level features. By the 5-layer time-delay neural networks (TDNNs), we can get the high-representation of frames. Then we compute the mean and standard deviation at the time dimension, and concatenate these two vectors, so-called statistics pooling layer. Fixed-dimensional utterance-level features are extracted through this operation. In the training stage, the network predicts the class of speaker by the last softmax layer with multi-class cross entropy loss. Besides, batch normalization and ReLU activation function are applied to all hidden layers.
\begin{equation}
    \mathscr{L}_{multi-class} = -\sum_{i=1}^{M}\frac{{\rm exp}^{w_{c_i}^Tx_i+b_{c_i}}}{\begin{matrix} \sum_{j}^N {\rm exp}^{w_j^Tx_i+b_j} \end{matrix}}
\end{equation}

In the evaluation stage, we let the last two layers as speaker embeddings (embedding A and embedding B respectively). Techniques such as PLDA or cosine similarity techniques are then applied to the extracted embeddings for scoring the trials.

\section{Joint learning of triplet distance and similarity network}
The entire architecture we proposed is as Fig \ref{fig:my_label}. Since embedding A of x-vector has better performance then embedding B, we add some constraints at the layer of embedding A. Triplet distance training directly controls the distance of two embeddings and the embedding similarity measurement network uses DNN to adjust the gap between embeddings.

\renewcommand\arraystretch{1.3}
\begin{table*}[h]
\begin{center}
\begin{threeparttable}
\caption{EER(\%) and DCF16 of SRE16 Evaluation Dataset}
\begin{tabular}{|c|p{120m}|c|c|c|c|c|c|c|}
\hline
\multicolumn{3}{|c|}{\centering System Description}
& \multicolumn{2}{c|}{\rule[-5pt]{0pt}{14pt} Pool}&\multicolumn{2}{c|}{\rule[-5pt]{0pt}{14pt} Tagalog}&\multicolumn{2}{c|}{\rule[-5pt]{0pt}{14pt} Cantonese}\\
\cline{1-9}
ID&\multicolumn{1}{|c|}{system}&triplet& \multicolumn{1}{c|}{\rule[-5pt]{0pt}{16pt}EER(\%)} & DCF16 & EER(\%) & DCF16 &EER(\%) &DCF16 \\
 \hline
1&\multicolumn{1}{|c|}{i-vector \cite{snyder2017deep}}& $\times$& 13.6 &0.711 &17.6&0.842&8.3&0.549 \\
2&\multicolumn{1}{|c|}{x-vector}& $\times$&8.65&0.679&12.50&0.829&4.80&0.460  \\

3&\multicolumn{1}{|c|}{x-vector, similarity net}&$\times$&8.42&{\bfseries0.633}&12.31&{\bfseries0.790}&4.58&{\bfseries0.424} \\
4&\multicolumn{1}{|c|}{x-vector, triplet distance}& embedding a&{\bfseries7.99}&0.686&{\bfseries11.70}&0.834&{\bfseries4.15}&{0.425}  \\

\cline{1-9}
5&\multicolumn{1}{|c|}{Joint Training, $\beta=0.1$, $\gamma=0.3$}& embedding a&{8.07}&{\bfseries0.617}&{11.98}&{\bfseries0.777}&{4.20}&  {0.412}\\
6&\multicolumn{1}{|c|}{Joint Training, $\beta=0.3$, $\gamma=0.1$}& embedding a&{\bfseries7.86}&{0.681}&{\bfseries11.51}&{0.834}&{\bfseries4.17}&{\bfseries0.411}  \\
7&\multicolumn{1}{|c|}{Joint Training,  $\beta=0.3$, $\gamma=0.1$, l2-norm}& embedding a&8.78&0.687&12.62&{0.833}&4.88&0.474  \\
8&\multicolumn{1}{|c|}{Joint Training, , $\beta=0.3$, $\gamma=0.1$}& embedding b&8.34&0.713&12.23&0.859&{4.30}&0.445  \\

\cline{1-9}
9&\multicolumn{1}{|c|}{fusion of system 2 and 4}& embedding a&7.39&0.644&10.87&0.807&3.82&0.397\\
10&\multicolumn{1}{|c|}{fusion of system 3 and 4}& embedding a&7.27&0.618&10.77&0.788&3.68&0.375\\

\cline{1-9}
\end{tabular}
\label{results}
\end{threeparttable}
\end{center}
\end{table*}

\subsection{Triplet Distance Training}
We random choice an utterance as the anchor, then choose an utterance from the same speaker as positive and an utterance from a different speaker as negative. Though cosine similarity is often used in back-end scoring, we find it can not converge if we use for restraining the triplet, thus we utilize Euclidean distance:
\begin{equation}
    d_{Euclidean} = \sqrt{ \sum_{i=1}^{N}(x_{1i}-x_{2i})^2}
\end{equation}
$x_1$and $x_2$ are two vectors, the triplet loss function is as follows:
\begin{equation}
   \mathscr{L}_{triplet} = {\lVert f({x}^{a}_{i})-f({x}^{p}_{i})\rVert}^{2}_{2}-{\lVert f({x}^{a}_{i})-f({x}^{n}_{i})\rVert}^{2}_{2}+a
\end{equation}
where $(x^{a}_{i},x^{p}_{i},x^{n}_{i})$ represents anchor, positive and negative embedding respectively, $a$ is the empirical margin and we set the value $a=0.8$. 

\subsection{Embedding Similarity Measurement Network}
Euclidean distances cannot fully reflect the similarity between the two embeddings. So we try to use another neural network to measure the similarity, called embedding similarity measurement network. We first obtain the embeddings though the x-vector network, cause these two embeddings have the same length, then we just concatenate the two embeddings as one sequence and feed it to the new network without delimiter in between. 
The similarity evaluation network consists of two Bidirectional Long Short-term Memory (BLSTM) layers with 1024 nodes and two fully connected layers with 512 nodes. Each hidden layer follows batch normalization and ReLU activation function.
Finally, the network gives the prediction of whether the two embeddings belong to one speaker. The loss function of this network is two-class cross-entropy loss:
\begin{equation}
    \mathscr{L}_{similarity} = -y {\rm log} \hat{y}-(1-y){\rm log}(1-\hat{y})
\end{equation}
where $y$ is the ground truth label and $\hat{y}$ is the predicted result.

As the proposed two losses have a different optimizing aim with the original multi-class cross entropy loss, we can jointly optimize these three losses by setting different weights.
\begin{equation}
    {\rm\mathscr{L}}_{total} = \alpha \mathscr{L}_{x-vector}+ \beta \mathscr{L}_{triplet} + \gamma \mathscr{L}_{similarity}
\end{equation}
where $\alpha$, $\beta$ and $\gamma$ are hyper-parameters and we set  $\alpha=1$, $\beta=0.1$ and $\gamma=0.3$ in our experiments.

\section{Experiments and Results}

\subsection{Dataset}
We use specific common corpora which are as follows for training: MIXER 6, 2010 NIST SRE and Follow-up Data, Switchboard 2 Phases 1, 2, and 3 as well as Switchboard Cellular. Data augmentation is an efficient way that makes the model more robust to the evaluation data. In our work, we add MUSAN dataset and room impulse responses (RIRs) data to the raw training data to expand the size and diversity of the training set by randomly choosing one among babble, music, noise and reverb. The RIRs and MUSAN are in 16k sampling rate and 16-bit precision. The process is applied using SRE16 recipe\footnote{https://github.com/kaldi-asr/kaldi/tree/master/egs/sre16/v2} of Kaldi speech recognition toolkit \cite{povey2011kaldi}.

After data augmentation, the training set consists of 5145 speakers and 211062 utterances.
We evaluate our system performance on NIST 2016 speaker recognition evaluations (SRE16), which consists of 802 speakers and 9294 utterances.

\subsection{Setup}
After voice activate detection (VAD), the augmentation audio is converted to 23-dimensional MFCC sequences with a frame-length of 25 ms, mean-normalized over a sliding window of 3 seconds. 
PyTorch toolkit is implemented for training the DNN. After getting the embedding, we use Kaldi PLDA in the evaluation stage as the backend for the experimental systems. Stochastic Gradient Descent (SGD) with weight decay=1e-8 and momentum=0.9 is used as the optimization method.
\subsection{Results and Analysis}

\begin{figure*}[h]
\centering
  {\includegraphics[scale=0.26]{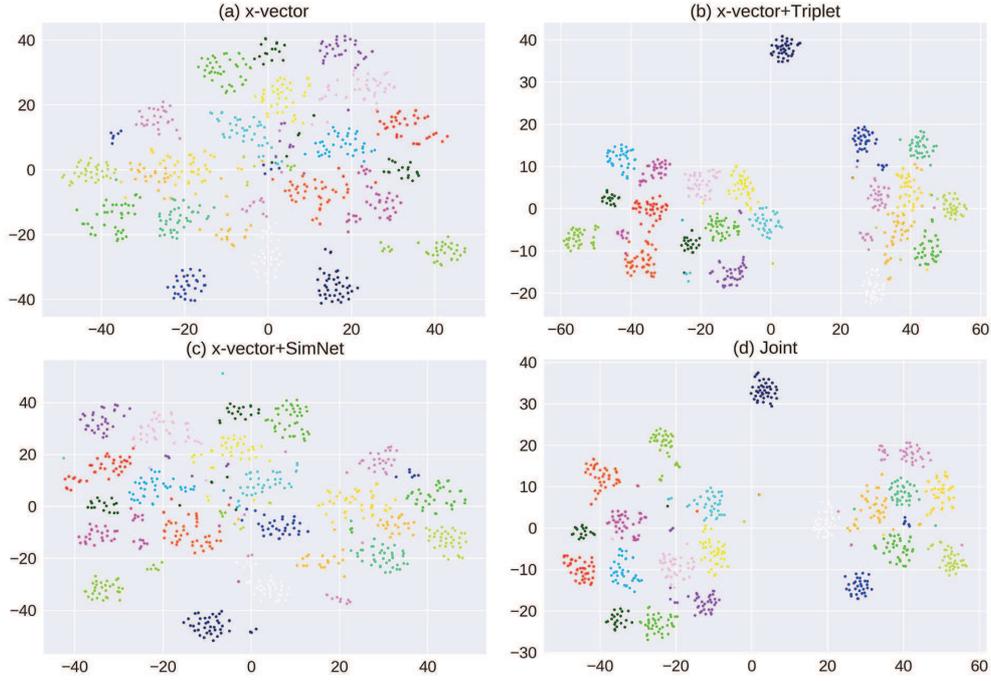}} 
\caption{visualization of different systems, plotted by the t-SNE, note the coordinate values of the longitudinal axis is different}
\label{vis}
\end{figure*}

We report the results in terms of EER and the SRE16 official evaluation metric DCF16, which is averaged from two operation points with $P_{Target} = 0.01$ and $P_{Target} = 0.005$ respectively. The results are listed in Table \ref{results}, in summary, systems with triplet distance training or embedding similarity measurement network outperform the baseline x-vector system that only trained with softmax loss.
We first investigate the influence of a single improvement. The two methods show their superiorities on different evaluation indicators.
More specifically, jointly training with embedding similarity measurement network (System 3) can reduce DCF obviously with a little effect of EER. At the same time, jointly training with triplet distance (System 4) remarkably decreases EER. 

Then we further explore the systems of three losses. 
The system 5\&6 evaluates the performances if we jointly train the baseline x-vector with the triplet distance and the embedding similarity measurement network. According to the results, we can observe that joint training achieves synthetically better performances compared with system 3\&4. And we can adjust the EER and DCF by finetune the hyperparameters $\beta$ and $\gamma$.
Adding l2-norm is a common way in triplet loss training task, but it seems not to make sense (system 7) in our scenario. We also test the same experiment at the embedding B layer (system 8), and we can conclude that embedding A is a better choice. 
Finally, we test two fusion systems (System 9 and 10) equally weighted of the PLDA by using Bosaris toolkit \cite{brummer2013bosaris}, the results can also prove that the similarity measurement network can improve the performance.

\begin{figure}
    \centering
    \includegraphics[scale=0.42]{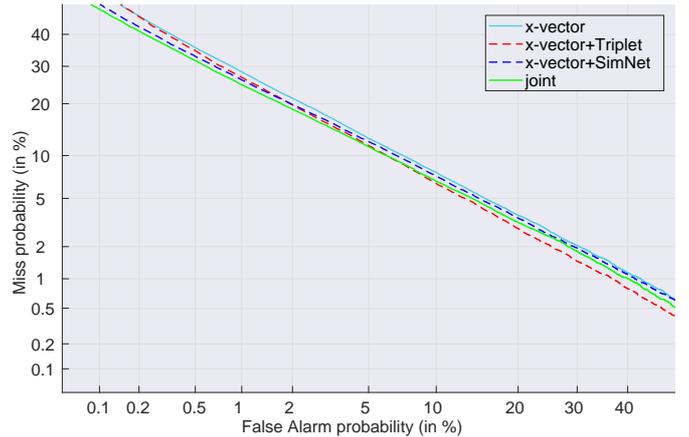}
    \caption{DET curve for the baseline and proposed systems when the results are pooled across Cantonese and Tagalog}
    \label{detcurve}
\end{figure}

\begin{figure}
    \centering
    \includegraphics[scale=0.41]{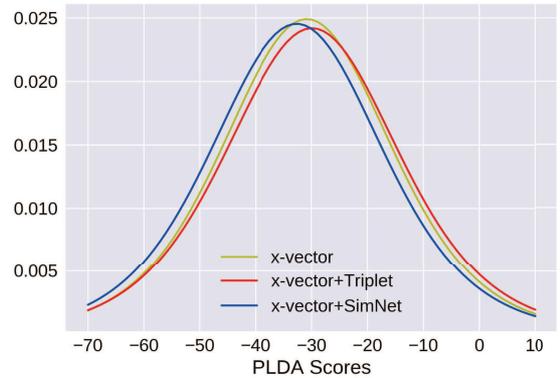}
    \caption{PLDA scores distribution of negative trials}
    \label{PLDA scores}
\end{figure}

Figure \ref{vis} shows the visualization of these systems. 
The observation is that the triplet distance loss shows reduced within-speaker variance, while the similarity measurement network shows its superior property on between-speaker distance.
Figure \ref{detcurve} plots the DET curve of system 2 to system 5. The results are corresponding to the table \ref{results}. The triplet distance can achieve better performance when the miss rate is very low, while the similarity measurement network performs better when the false alarm rate is low, and the performance of joint training is slightly better than the triplet distance. 
It is also observed that non-target scores distribution of system 3 is more negative than other systems in Figure \ref{PLDA scores}, which is consistent with the results in Table \ref{results} and Figure \ref{detcurve}.
Combining these factors, it is proved that the embedding similarity measurement network can truly improve the performance.

\section{Conclusions}
In this work, we focus on a text-independment speaker verification task and present two improvements based on triplet. We firstly implement triplet distance training and reduce EER. Getting inspiration of NLP, we design an embedding similarity measurement network for classifying whether the two embeddings belong to one speaker in the training stage, meaning we add another constraint on the embeddings. The performance shows that adding the new network can get a definite decrease of DCF while lowing the EER. 
Finally we get a better performance by jointly training with the three losses.
The results shows benefit from our embedding similarity measurement network, therefore, the proposed methods can be applied to other end-to-end verification systems.

\bibliographystyle{IEEEtran}

\bibliography{mybib}

\end{document}